\documentclass[fdp,a4paper,fleqn
]{w-art}
\usepackage{times,cite,w-thm}
\usepackage{textcomp}
\theoremstyle{plain}

\theoremstyle{definition}

\usepackage[]{graphicx}
\begin{document}
\DOIsuffix{theDOIsuffix}
\pagespan{1}{}


\title[Atom-Photon]{Protocols and Techniques for a Scalable Atom--Photon Quantum Network}
\author{L. Luo \footnote{Corresponding author\quad
E-mail:~\textsf{leluo@umd.edu}}}
\author{D. Hayes}
\author{T.A. Manning}
\author{D.N. Matsukevich}
\author{P. Maunz}
\author{S. Olmschenk}
\author{J.D. Sterk}
\author{C. Monroe}
\address{Joint Quantum Institute, University of Maryland Department of Physics and
National Institute of Standards and Technology, College Park, MD
20742}
\begin{abstract}
Quantum networks based on atomic qubits and scattered photons
provide a promising way to build a large-scale quantum information
processor. We review quantum protocols for generating entanglement
and operating gates between two distant atomic qubits, which can be
used for constructing scalable atom--photon quantum networks. We
emphasize the crucial role of collecting light from atomic qubits
for large-scale networking and describe two techniques to enhance
light collection using reflective optics or optical cavities. A
brief survey of some applications for scalable and efficient
atom--photon networks is also provided.
\end{abstract}


\maketitle






\section{Introduction}

Scalable quantum information processing has been proposed in many
physical systems~\cite{DiVincenzo:2000,Monroe:Review:2008}. There
are two types of quantum bits (qubits) currently under
investigation: material and photonic qubits. Qubits stored in atoms
or other material systems can behave as good quantum memories, but
they are generally difficult to transport over large distances. On
the other hand, photonic qubits are appropriate for quantum
communication over distance, yet are difficult to store. It is thus
natural to consider large--scale quantum networks that reap the
benefits of both types of quantum platforms.

Trapped atomic ions are among the most attractive candidates for
quantum memory, owing to their long storage and coherence
times~\cite{Blatt:Review:2008,Haffner:Review:2008}. The traditional
approach to entangle multiple trapped ions relies on a direct
Coulomb interaction between ions in close proximity
~\cite{leibfried:2003b,sk:2003,Haljan:2005,Home:2006}. Scaling to
larger numbers of qubits can then proceed by shuttling ions between
multiple trapping zones in complex trap structures
~\cite{kielpinski:2002}.

\begin{figure}[tb]
\begin{center}
\includegraphics[width= 3.0in]{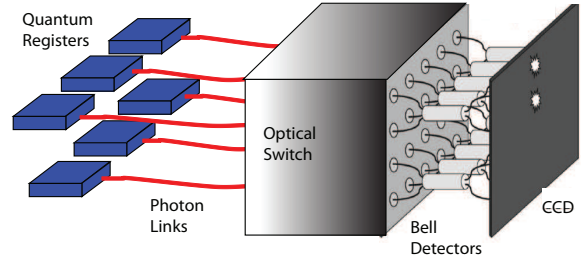}
\caption{Concept of a distributed atom--photon network for a
large-scale quantum information processor. Quantum registers are ion
traps containing singular or locally interacting atomic qubits (blue
boxes), which are linked by photons (red lines) through the use of
an optical switch and optical Bell-state detectors.} \label{fig:net}
\end{center}
\end{figure}

A higher level architecture for entangling an arbitrary number of
atomic qubits is an atom--photon network
~\cite{duan:2004,Moehring:Atomphoton:2007,Duan:Review:2008,Kimble:internet:2008,Kim:2009},
in which singular or locally interacting material qubits form the
quantum registers of a distributed quantum network as depicted in
Fig.~\ref{fig:net}. The modular structure of this network relies on
entangling atomic qubits with photonic qubits in order to establish
entanglement between the quantum registers. Here, we concentrate on
probabilistic links where the atom--photon entanglement process
succeeds with a low probability, yet this probabilistic process
still can be used to to generate arbitrary-size quantum
networks~\cite{duan:2005}.

Central to the idea of a probabilistic atom--photon network is the
heralded entanglement process between remote nodes, which includes
the following two steps. First, the scattering of photons
establishes entanglement between atomic and photonic qubits. Second,
the interference and detection of the scattered photons from
multiple atomic qubits project the remote atomic qubits into an
entangled state. The success probability for the first step is
\begin{equation}
P_{ap} = p_{e}\,p_{c}\,p_{t},
    \label{eqn:ProbQphot}
\end{equation}
where $p_{e}$ is the probability of generating the desired entangled
atom--photon pair, $p_{c}$ is the efficiency of collecting the
emitted photon by the optical system, and $p_{t}$ is the
transmission efficiency of the entire optical system, including
fiber coupling. The second step of the heralded entanglement adds
other factors to the success probability, including the loss of
certain photonic states at the Bell state
detector~\cite{braunstein:1995} and the quantum efficiency of the
photon detectors. Experiments demonstrating entanglement between two
remote
ions~\cite{Moehring:Entangle:2007,Matsukevich:Bell:2008,Olmschenk:Tele:
2009,Maunz:Gate:2009} have clearly shown that the light collection
efficiency $p_c$ is the leading limiting factor for the overall
success probability. Improvements to the light collection efficiency
from trapped ions therefore play a crucial role for creating a
scalable and efficient atom--photon quantum network.

In this paper, we review protocols and techniques suitable for
constructing a scalable atom--photon quantum network. We first
provide a detailed description of methods to create entangled
atom--photon pairs with different types of photonic qubits
(Sec.~\ref{sec:protocol}), as well as protocols for generating
heralded entanglement and operating quantum gates between two remote
atomic qubits (Sec.~\ref{sec:entanglement}). Where applicable, we
will provide examples with ${}^{171}\mathrm{Yb}^{+}$ ions as quantum
memories. To apply these schemes for building large--scale
atom--photon networks, we discuss specific protocols and techniques
relevant to trapped ions, including using photon emission from
multiple ions to entangle a linear crystal of ions
(Sec.~\ref{sec:multiion}) and improving light collection from
trapped ions by reflective optics or an optical cavity
(Sec.~\ref{sec:MethodsLC}).  We also present a brief outlook of a
scalable atom--photon network for quantum information processing
(Sec.~\ref{sec:outlook}).


\section{Protocols for Generating Atom--Photon Entanglement\label{sec:protocol}}
Entangled atom--photon pairs can be generated from a wide range of
systems, including trapped ions, neutral atoms, and
atomic ensembles~\cite{Duan:Review:2008}. The quantum protocols used to
produce these entangled pairs are general and not limited to
the specific systems. In this section, we first review these general
protocols, and then use trapped ${}^{171}\mathrm{Yb}^{+}$ ions as an
example to show detailed schemes of generating both ultraviolet
and infrared photonic qubits.

\subsection{Types of Photonic Qubits\label{sec:pairs}}
When a laser pulse hits an atom, photons can be scattered through
either a resonant or off-resonant process. In both processes the
atom can be transferred from the initial state to the multiple final
states through different scattering channels.
If an atom has two distinct scattering channels correlated
with two orthogonal states of the scattered photon, an entangled
atom--photon pair is produced as described by
\begin{equation}
|\Psi_{ap}\rangle= c_\uparrow\mid\uparrow\rangle|P_\uparrow\rangle +
c_\downarrow\mid\downarrow\rangle|P_\downarrow\rangle,
\label{eq:atom-photon}
\end{equation}
where $\mid\uparrow\rangle$ and $\mid\downarrow\rangle$ represent
the atomic qubit state, and $|P_\uparrow\rangle$ and
$|P_\downarrow\rangle$ are the orthogonal states of the photonic
qubit. The values of the probability amplitudes, $c_\uparrow$ and
$c_\downarrow$, depend on the specific scheme used to generate the
entangled pairs.

The physical properties of photons that are most often used to
encode the photonic qubits are (i) existence of the photon(s)
(number qubits), (ii) polarization, (iii) frequency, and (iv) the
emission time (time-bin qubits).

\begin{figure}[tb]
\includegraphics[width=5.5 in]{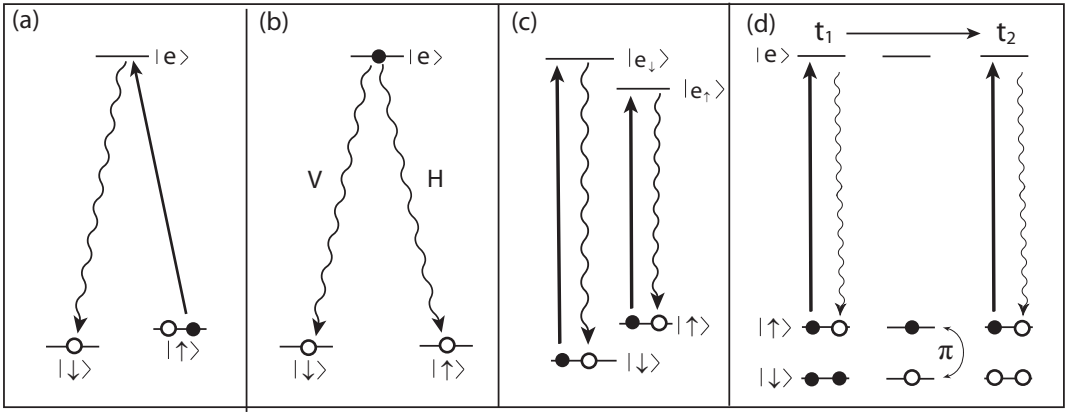}
\caption{Protocols for generating entangled atom--photon pairs.
Black dots represent initial atomic states, and white dots represent
final atomic states. The solid straight lines represent photons in
the laser beam. The wiggly lines represent photons scattered or
emitted from the atom. The atomic qubit states are denoted by
$\mid\uparrow\rangle$ and $\mid\downarrow\rangle$. \textbf{(a)
Number Qubit.} A photon is scattered from a laser pulse,
transferring the atom from the initial state to the final state with
a certain probability. The existence of the scattered photon is
entangled with the atomic qubits. \textbf{(b) Polarization Qubit.}
An excited atom with two effective decay channels spontaneously
emits a photon whose polarization is correlated with the final
atomic state, thus generating entanglement between the polarization
of the emitted photon and the atomic qubits. \textbf{(c) Frequency
Qubit.} An atom in an arbitrary qubit state is coherently
transferred by a broadband pulse to excited states. The selection
rules ensure that the population in each excited state decay back to
its original state, generally entangling the frequency of
spontaneously emitted photons with the atomic qubit. \textbf{(d)
Time-bin Qubit.} An atom is initially in an arbitrary qubit state. A
laser pulse resonant with the $\mid\uparrow\rangle \leftrightarrow
\mid e \rangle$ transition is applied to the atom. After the
population in the excited state spontaneously decays back to the
initial state, a $\pi$ pulse rotates the qubit. A second identical
laser pulse is subsequently applied, followed by another spontaneous
decay. The emission time of the photon defines the photonic qubit,
which is generally entangled with the atomic qubit.}
\label{fig:photonicqubits}
\end{figure}

\textbf{Number Qubits.} Suppose a three-level atom with a `$\Lambda$
configuration' is initially prepared in one of its ground states,
$\mid\uparrow\rangle$, as shown in Fig.~\ref{fig:photonicqubits}(a).
Photons in a laser pulse are scattered by the atom, transferring the
atom from $\mid\uparrow\rangle$ to $\mid\downarrow\rangle$ with
certain probability $p_e<1$. After the scattering, the final
atom--photon state is described by
\begin{equation}
|\Psi_{ap}\rangle= \sqrt{1-p_e}\mid\uparrow\rangle|0\rangle +
\sqrt{p_e}\mid\downarrow\rangle|1\rangle\,\,,
\end{equation}
conforming to Eq.~\ref{eq:atom-photon}. Here $|0\rangle$ and
$|1\rangle$ denote the vacuum state and the one photon state
generated by the scattering process.

\textbf{Polarization Qubits.} Polarization qubits can be realized by
the scheme shown in Fig.~\ref{fig:photonicqubits}(b).  An atom,
initially prepared in the excited state $|e\rangle$, spontaneously
decays to the ground states $\mid\uparrow\rangle$ and
$\mid\downarrow\rangle$ with probability $p_\uparrow$ and
$p_\downarrow$, respectively. Each decay channel is associated with
a polarization component of the emitted photon. If these two
polarization components, denoted by $H$ and $V$, are orthogonal, the
atomic qubit stored in the $\mid\uparrow\rangle$ and
$\mid\downarrow\rangle$ states is entangled with the polarization of
the emitted photon. The overall atom-photon state is described by
\begin{equation}
|\Psi_{ap}\rangle=
\sqrt{p_\uparrow}\mid\uparrow\rangle|1_H0_V\rangle +
\sqrt{p_\downarrow}\mid\downarrow\rangle|0_H1_V\rangle\,\,.
\end{equation}
conforming to Eq.~\ref{eq:atom-photon}. Here $|1_H0_V\rangle$ and
$|0_H1_V\rangle$ denote the $H$ and $V$ polarization states for a
single photon.

\textbf{Frequency Qubits.}  One example of generating frequency
qubits is shown in Fig.~\ref{fig:photonicqubits}(c), where both the
ground state and the excited state have two non-degenerate energy
levels. Frequency qubits can also be generated by addressing only
one energy level in the excited state, but this configuration does
not support a quantum gate operation ~\cite{Moehring:Entangle:2007}.
In the case shown in Fig.~\ref{fig:photonicqubits}(c), the four
energy levels are chosen so that selection rules only permit
transitions from
$\mid\uparrow\rangle\leftrightarrow|e_\uparrow\rangle$ and
$\mid\downarrow\rangle\leftrightarrow|e_\downarrow\rangle$. The atom
is initially prepared in an arbitrary superposition state,
$\alpha\mid\uparrow\rangle + \beta\mid\downarrow\rangle$. A laser
pulse, whose bandwidth covers both the
$\mid\uparrow\rangle\leftrightarrow|e_\uparrow\rangle$
$\mid\downarrow\rangle\leftrightarrow|e_\downarrow\rangle$
transitions, coherently excites the atom to the
$\alpha|e_\uparrow\rangle + \beta|e_\downarrow\rangle$ state. The
population in each excited state then decays back to its initial
ground state. Because the frequency of the spontaneously emitted
photon is uniquely correlated with the originally prepared quantum
state, this process results an entangled atom--photon pair given by
\begin{equation}
|\Psi_{ap}\rangle= \alpha\mid\uparrow\rangle|1_r0_b\rangle +
\beta\mid\downarrow\rangle|0_r1_b\rangle\,\,,
\end{equation}
conforming to Eq.~\ref{eq:atom-photon}. Here $|1_r0_b\rangle$ and
$|0_r1_b\rangle$ denote the one-photon states for two different
frequency components, where $r$ is the red, or low frequency photon,
and $b$ is the blue, or high frequency photon. These frequency qubit
states are resolved when $\omega_b-\omega_r\gg\Gamma$, where
$\omega_{b}$ and $\omega_{r}$ are the two photon frequencies and
$\Gamma$ is the linewidth of the transitions. We note that the final
entangled state preserves the quantum information in the initial
qubit state, thus admitting quantum gate operations.

\textbf{Time-bin Qubits.} The time at which a photon is emitted
can also be used to encode photonic
qubits~\cite{Brendel:timebin:1999,barrett:2005}. This type of
photonic qubit is defined by a superposition of states in which a
photon exists within a `time bin' centered either at time $t_1$ or
$t_2$. The time-bin states are resolved when
$e^{-\Gamma~|t_2 - t_1|}\ll1$. As shown in Fig.~\ref{fig:photonicqubits}(d), an
atom is initially prepared in a superposition state of
$\alpha\mid\uparrow\rangle + \beta\mid\downarrow\rangle$. The
excited energy level $|e\rangle$ is chosen so that the selection
rules only permit the transition from
$\mid\uparrow\rangle\leftrightarrow|e\rangle$. At time $t_1$, a single
frequency laser pulse resonant with the
$\mid\uparrow\rangle\leftrightarrow|e\rangle$ transition excites the
population to the $\mid\uparrow\rangle$ state. After the population
in the excited state decays, the atom--photon state is
$\alpha\mid\uparrow\rangle|1_{t_1}\rangle+\beta\mid\downarrow\rangle|0_{t_1}\rangle$,
where $|1_{t_1}\rangle$ and $|0_{t_1}\rangle$ represent the
one-photon state and the vacuum state at time $t_1$. A $\pi$ pulse
is then applied to rotate the atomic qubit, changing the atom--photon
state to
$-\alpha\mid\downarrow\rangle|1_{t_1}\rangle+\beta\mid\uparrow\rangle|0_{t_1}\rangle$.
At time $t_2$, a second resonant laser pulse is applied to
again excite the population in the $\mid\uparrow\rangle$ state.
Following the spontaneous decay, the final entangled atom--photon
state is given by
\begin{equation}
|\Psi_{ap}\rangle= \beta\mid\uparrow\rangle|0_{t_1}1_{t_2}\rangle
-\alpha\mid\downarrow\rangle|1_{t_1}0_{t_2}\rangle\,\,,
\end{equation}
conforming to Eq. \ref{eq:atom-photon}. Here
$|0_{t_1}1_{t_2}\rangle$ and $|1_{t_1}0_{t_2}\rangle$ represent the
states of a single photon at time $t_2$ and $t_1$.

\subsection{Photonic Qubits for ${}^{171}\mathrm{Yb}^{+}$ Ions \label{sec:ybqubits}}
All the atom--photon entanglement schemes discussed above can be
realized through resonant scattering processes with laser-cooled,
trapped ${}^{171}\mathrm{Yb}^{+}$ ions~\cite{olmschenk:2007}.
Table~\ref{tbl:qubits} details how the four types of photonic qubits
illustrated in Fig.~\ref{fig:photonicqubits} can be created using
the ${}^{2}S_{1/2}\leftrightarrow{}^{2}P_{1/2}$ transition of
${}^{171}\mathrm{Yb}^{+}$ ions in the ultraviolet (UV) at $370$~nm.
Polarization and frequency photonic qubits have been realized
experimentally in
Ref.~\cite{Moehring:Entangle:2007,Matsukevich:Bell:2008,Olmschenk:Tele:2009,Maunz:Gate:2009}.

\begin{figure}[tb]
\begin{center}
\includegraphics[width= 3.0in]{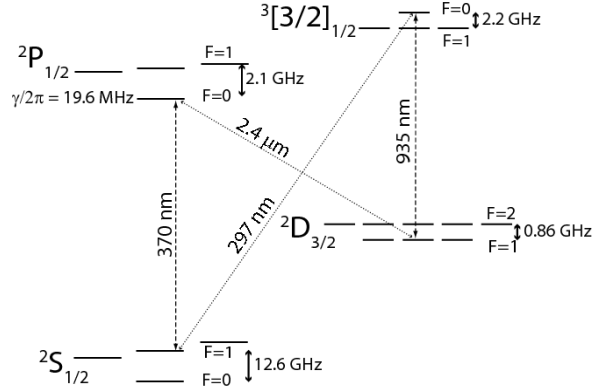}
\caption{Relevant energy levels for ${}^{171}\mathrm{Yb}^{+}$. The
${}^{2}S_{1/2}\leftrightarrow{}^{2}P_{1/2}$ transition is driven by
light at 370~nm. The ion decays from ${}^{2}P_{1/2}$ to the
${}^{2}D_{3/2}$ state with probability of 0.005. A 935~nm
laser is used to pump the ion out of this state through the
$^{3}[3/2]_{1/2}$ level. The ${}^{3}P_{3/2}$ state is above the
${}^{2}P_{1/2}$ state, but is not shown in the figure. }
\label{fig:Ybatomic}
\end{center}
\end{figure}

\begin{table}[ht]
\begin{center}
\scalebox{1.0} {\begin{tabular}{|l|c|c|}
  \hline
  Photonic Qubit Type& Ground State & Excited State \\
  \hline \hline
  Number Qubit & $\mid\uparrow\rangle={}^{2}S_{1/2} |0,0\rangle$ & $|e\rangle={}^{2}P_{1/2}|1,-1\rangle$ \\
               & $\mid\downarrow\rangle={}^{2}S_{1/2} |1,-1\rangle$ & \\
  \hline
  Polarization Qubits &$\mid\uparrow\rangle={}^{2}S_{1/2} |1,1\rangle$  & $|e\rangle={}^{2}P_{1/2}|0,0\rangle$ \\
                      &$\mid\downarrow\rangle={}^{2}S_{1/2} |1,-1\rangle$ & \\
  \hline
  Frequency Qubits & $\mid\uparrow\rangle={}^{2}S_{1/2} |1,0\rangle$  & $|e_\uparrow\rangle={}^{2}P_{1/2}|0,0\rangle$ \\
                   & $\mid\downarrow\rangle={}^{2}S_{1/2} |0,0\rangle$ & $|e_\downarrow\rangle={}^{2}P_{1/2}|1,0\rangle$ \\
  \hline
  Time-bin Qubits  & $\mid\uparrow\rangle={}^{2}S_{1/2} |1,0\rangle$ & $\mid e\rangle={}^{2}P_{1/2}|0,0\rangle$ \\
                   & $\mid\downarrow\rangle={}^{2}S_{1/2} |0,0\rangle$ & \\
  \hline
\end{tabular}}
\caption{The energy levels for generating ultraviolet photonic
qubits from the ${}^{2}P_{1/2}\leftrightarrow{}^{2}S_{1/2}$
transition, where the Zeeman levels are denoted by $|F,m_F\rangle$.}
\label{tbl:qubits}
\end{center}
\end{table}

\begin{figure}[tb]
\begin{center}
\includegraphics[width= 5.5in]{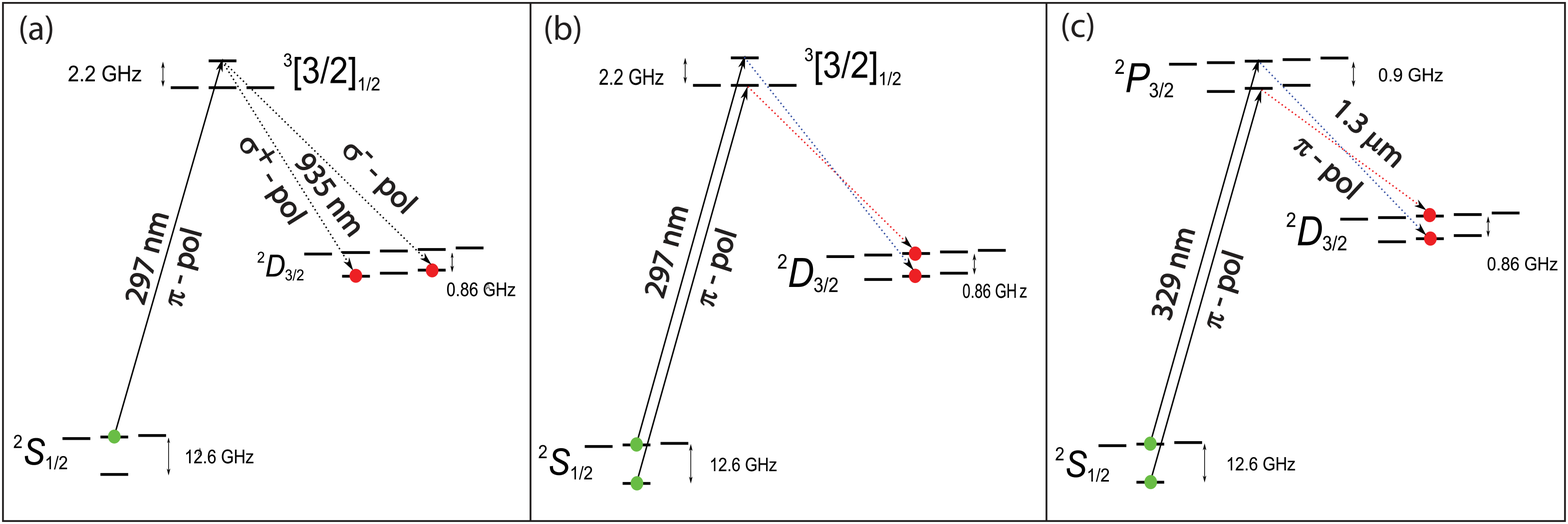}
\caption{Generation of infrared photonic qubits entangled with
${}^{171}\mathrm{Yb}^{+}$ ions. (a) Generating polarization photonic
qubits at 935~nm. An atom is first prepared in the ${}^{2}S_{1/2}
\vert F=1,m_{F}=0 \rangle$ state. A $\pi$-polarized pulse at
297.1~nm transfers the atom to the ${}^{3}[3/2]_{1/2} \vert
F=0,m_{F}=0 \rangle$ state, while transitions to the
${}^{3}[3/2]_{1/2} \vert F=1,m_{F}=0\rangle$ level are forbidden.
The atom can spontaneously decay to the ${}^{2}D_{3/2} \vert F=1
\rangle$ level and emit a 935~nm photon, resulting the entanglement
between the polarization qubits of the 935~nm photon and the atomic
qubits in the hyperfine levels of the ${}^{2}D_{3/2} \vert F=1
\rangle$ state. (b) Generating frequency photonic qubits at 935~nm.
An atom is initialized to a superposition state including the
$|F=0,m_{F}=0 \rangle$ and $|F=1,m_{F}=0 \rangle$ hyperfine levels
in the ${}^{2}S_{1/2}$ state. Then a $\pi$-polarized pulse at 297~nm
coherently transfers the population from ${}^{2}S_{1/2}$ to
${}^{3}[3/2]_{1/2}$. The selection rules only permit ${}^{2}S_{1/2}
\vert F=0,m_{F}=0 \rangle\leftrightarrow{}^{3}[3/2]_{1/2} \vert
F=1,m_{F}=0 \rangle$ and ${}^{2}S_{1/2} \vert F=1,m_{F}=0
\rangle\leftrightarrow{}^{3}[3/2]_{1/2} \vert F=0,m_{F}=0 \rangle$.
Then the ${}^{3}[3/2]_{1/2}$ levels can decay to ${}^{2}D_{3/2}$ by
spontaneously emitting a 935~nm photon. By only collecting
$\pi$-polarized photons the frequency of the emitted photon is
entangled with the hyperfine state of ${}^{2}D_{3/2}$. (c)
Generating frequency photonic qubits at 1.3~$\hbox{\textmu}$m. An
atom is initially prepared ${}^{2}S_{1/2}$ and then excited to
${}^{2}P_{3/2}$ by a laser pulse at 329~nm. Decay from
${}^{2}P_{3/2}$ to ${}^{2}D_{3/2}$ results in the emission of a
1.3~$\hbox{\textmu}$m photon, producing the same atomic qubit as in
(b), but with a different frequency photonic qubit of wavelength
1.3~$\hbox{\textmu}$m.}\label{fig:infrared_entangle}
\end{center}
\end{figure}

We note that the ${}^{171}\mathrm{Yb}^{+}$ system also supports infrared (IR)
photonic qubits, which may be useful for long-distance quantum communication.
Moreover, access to additional
optical frequencies may facilitate entanglement between disparate
optically active systems (\emph{e.g.} between trapped ions and
quantum dots). IR photonic qubits can be generated by either the
${}^{3}[3/2]_{1/2} \leftrightarrow {}^{2}D_{3/2}$ transition
(935~nm) or the ${}^{2}P_{3/2} \leftrightarrow {}^{2}D_{3/2}$
transition (1.3~$\hbox{\textmu}$m), as shown in
Fig.~\ref{fig:infrared_entangle} for polarization and
frequency photonic qubits.

The branching ratio of the ${}^{3}[3/2]_{1/2}$ level between
${}^{2}S_{1/2}$ and ${}^{2}D_{3/2}$ has been calculated to be about
55:1 ~\cite{biemont:1998}, which decreases the probability of
generating a 935~nm photon and thereby reduces the overall success
probability of the entanglement protocols discussed below. Two
potential experimental problems arise from the small branching
ratio. First, a higher fraction of the detection events will be dark
counts due to the additional detector integration time. However,
there is a way to ``veto'' these extra dark counts. After a 935~nm
photon detection event, a $\pi$ pulse at the ${}^{2}D_{3/2}$
hyperfine splitting can be used to coherently shelve the qubit state
populations to the ${}^{2}D_{3/2} \vert F=2 \rangle$ manifold. If
the fluorescence detection procedure detailed in
\cite{olmschenk:2007} is now performed, in which population of
${}^{2}D_{3/2} \vert F=2 \rangle$ is undisturbed, no 370~nm photons
should be detected. The detection of 370~nm photons during this
interval would indicate that the 935~nm photon ``detection'' was in
fact a dark count and should be discarded. The atomic population
transferred from ${}^{2}D_{3/2} \vert F=1 \rangle$ to ${}^{2}D_{3/2}
\vert F=2 \rangle$ can then be returned by a second microwave $\pi$
pulse. The second problem posed by the small branching ratio is that
it is difficult to implement state detection for atomic qubits by
using the 935~nm infrared transition directly. Instead, the
${}^{2}D_{3/2}$ states can be mapped to the ${}^{2}S_{1/2}$ states
so that the aforementioned UV photon fluorescence detection can be
used to measure the atomic qubits. For example, to detect the atomic
states described in Fig.~\ref{fig:infrared_entangle}(a), the
population in the ${}^{2}D_{3/2}\vert F=1,m_{F}=1 \rangle$ state
could be transferred to the ${}^{2}D_{3/2} \vert F=2 \rangle$
manifold by a resonant microwave pulse at the hyperfine splitting of
0.86~GHz. Light from a 935~nm laser could transfer the population
from the ${}^{2}D_{3/2} \vert F=1 \rangle$ manifold to
${}^{2}S_{1/2} \vert F=1 \rangle$ (Fig.~\ref{fig:Ybatomic}). The
state of the atom is then determined by resonantly driving the
${}^{2}S_{1/2} \vert F=1 \rangle \leftrightarrow {}^{2}P_{1/2} \vert
F=0 \rangle$ transition and detecting the fluorescence, indicating
the atom was originally in the ${}^{2}D_{3/2}\vert F=1,m_{F}=-1
\rangle$ state.

Finally, we consider the 1.3~$\hbox{\textmu}$m photons generated
from the ${}^{2}P_{3/2}$ to ${}^{2}D_{3/2}$ transition. The
branching ratio from ${}^{2}P_{3/2}$ to ${}^{2}S_{1/2}$ versus
${}^{2}D_{3/2}$ is about 475:1 ~\cite{biemont:1998}. While the
1.3~$\hbox{\textmu}$m wavelength is more amenable to long distance
transmission, the decrease in the protocol success probability is
even more drastic. In addition, ${}^{2}P_{3/2}$ can also decay to
${}^{2}D_{5/2}$, which can subsequently decay to the long-lived
${}^{2}F_{7/2}$ state. Depopulating these additional metastable
states could limit the repetition rate of the experiment.


\section{Protocols for Generating Remote Atom--Atom Entanglement \label{sec:entanglement}}

\begin{figure}[tb]
\begin{center}
\includegraphics[width= 5 in]{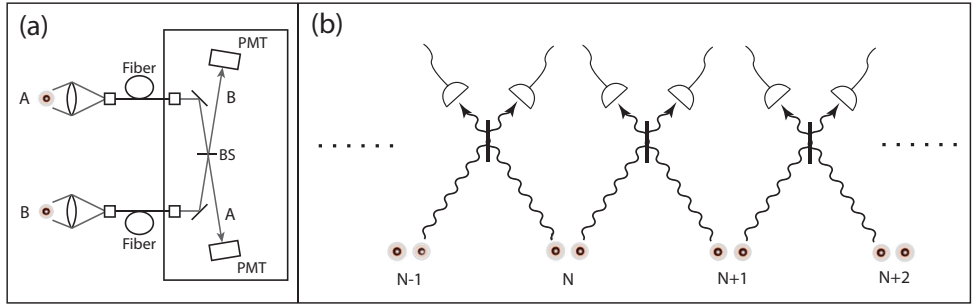}
\caption{ \textbf{(a)} The experimental setup for the interference
effect of individual photons emitted by two ions. The 50:50
beamsplitter (BS) ensures that the origins of detected photons are
unknown. In type I heralded entanglement, only one spontaneously
emitted photon is detected following the excitation. Type II
heralded entanglement involves two photons interfering on the
beamsplitter. \textbf{(b)} Conceptual sketch of an atom--photon
quantum repeater, in which locally-interacting atomic qubits are
entangled with remote atomic qubits via a heralded entanglement
protocol and local gate operations. } \label{fig:photoninter}
\end{center}
\end{figure}

The entangled atom--photon pairs described in the last section can
be used to entangle two remote non-interacting atomic qubits. The
key component of entangling these atomic qubits is the interference
of the photons on a 50:50 beamsplitter as shown in
Fig.~\ref{fig:photoninter}(a). In contrast to post-selected
entanglement schemes, in which measurement
of the entangled qubits destroys the entanglement, the detection of
photons after the beamsplitter destroys only the photonic system,
which can herald the projection of the atomic states into an
entangled state. This heralded entanglement technique generates
useful entangled atomic pairs that serve as the chief resource for
constructing a photon-mediated quantum network.

There are two types of heralded entanglement, distinguished by
whether only one photon is emitted by the two atoms (type I) or one
photon is emitted by \emph{each} of the two atoms (type
II)~\cite{duan:2004}. For a single photon input
state~\cite{mandel:1999,Moehring:Atomphoton:2007}, the output of a
50:50 beamsplitter is given by
\begin{eqnarray}
|0\rangle_A |1\rangle_B & \underrightarrow{BS}&\frac{|0\rangle_A
|1\rangle_B+|1\rangle_A |0\rangle_B}{\sqrt{2}}\,\,,\nonumber\\
|1\rangle_A |0\rangle_B & \underrightarrow{BS}&\frac{-|0\rangle_A
|1\rangle_B+|1\rangle_A |0\rangle_B}{\sqrt{2}}\,\,. \label{eq:onep}
\end{eqnarray}
As shown in Fig.~\ref{fig:photoninter}(a),\,$|i\rangle_A|j\rangle_B$
denotes the number of photons $i$ and $j$ in the two spatial modes
$A$ and $B$ entering or emerging from the beamsplitter, having the
same frequency and polarization. For a two-photon input
state~\cite{mandel:1999,Moehring:Atomphoton:2007}, the interference
generates the output states
\begin{eqnarray}
|0\rangle_A|2\rangle_B&\underrightarrow{BS}&\frac{|0\rangle_A|2\rangle_B}{2}+
\frac{|1\rangle_A|1\rangle_B}{\sqrt{2}}+ \frac{|2\rangle_A|0\rangle_B}{2}\,\,,\nonumber\\
|1\rangle_A|1\rangle_B&\underrightarrow{BS}&\frac{(-|0\rangle_A|2\rangle_B+|2\rangle_A|0\rangle_B)}{\sqrt{2}}\,\,,\nonumber\\
|2\rangle_A|0\rangle_B&\underrightarrow{BS}&\frac{|0\rangle_A|2\rangle_B}{2}-
\frac{|1\rangle_A|1\rangle_B}{\sqrt{2}}+\frac{|2\rangle_A|0\rangle_B}{2}\,\,.
\label{eq:twop}
\end{eqnarray}


\subsection{Type I Heralded Entanglement}
To implement the type I heralded entanglement protocol, laser pulses are applied
to two atoms $A$ and $B$ so that $p_e \ll 1$, as shown in Fig.~\ref{fig:photonicqubits}(a).
The scattering process yields two entangled atom--photon pairs with states
$|\Psi_{ap}\rangle=\sqrt{1-p_e}\mid\uparrow\rangle|0\rangle +
\sqrt{p_e}\mid\downarrow\rangle|1\rangle$. The state of the two
atom--photon pairs $A$ and $B$ is then
\begin{eqnarray}
|\Psi_{apap}\rangle &=&|\Psi_{ap}\rangle_A\otimes|\Psi_{ap}\rangle_B\nonumber\\
&\approx&\mid\uparrow\rangle_A\mid\uparrow\rangle_B|0\rangle_A|0\rangle_B+\sqrt{p_e}\big(\mid\uparrow\rangle_A\mid\downarrow\rangle_B|0\rangle_A|1\rangle_B+e^{i\phi}|\downarrow\rangle_A|\uparrow\rangle_B|1\rangle_A|0\rangle_B\big),
\end{eqnarray}
where terms of order $p_e$ and higher are ignored. The relative
phase $\phi={\Delta k\,\Delta x}$, where $\Delta k$ is the
wavevector difference between the excitation laser and the collected
photons, and $\Delta x$ is the optical path lengths difference from
the atoms to the beamsplitter. After the beamsplitter, the output
state containing one photon is
\begin{equation}
|\Psi_{apap}\rangle =\big(\mid\uparrow\rangle_A
\mid\downarrow\rangle_B - e^{i\phi} \mid\downarrow\rangle_A
\mid\uparrow\rangle_B\big)|0\rangle_A|1\rangle_B+\big(\mid\uparrow\rangle_A
\mid\downarrow\rangle_B + e^{i\phi}\mid\downarrow\rangle_A
\mid\uparrow\rangle_B\big)|1\rangle_A|0\rangle_B.
\end{equation}
This result indicates that once the two PMTs detect one photon, the atom--atom state is projected into one of the following states
\begin{equation}
|\Psi_{aa}\rangle=\mid\uparrow\rangle_A \mid\downarrow\rangle_B \pm
e^{i\phi} \mid\downarrow\rangle_A \mid\uparrow\rangle_B,
\label{eq:type1ent}
\end{equation}
where the sign is determined by which PMT detects the photon.

The success probability of type I entanglement is
$P_I=2\,P_{ap}\,\eta_{det}$, where $P_{ap}$ is the success
probability of obtaining a useful atom--photon entangled pair
(Eq.~\ref{eqn:ProbQphot}) and $\eta_{det}$ is the photon detection
efficiency. According to the above analysis, there is a
small probability of order $p_e^2$ that both atoms
scatter photons. If only one photon is detected, there is still a
probability $p_e$ that another photon was emitted but
not detected, representing a limit on the fidelity of this entanglement procedure.

Type I entanglement relies on interferometric stability of the
optical paths, so fluctuations in the phase $\phi = \Delta k \Delta
x$ must be kept small. An important source of decoherence is the
atomic recoil from the absorption and emission of a photon,
indicating which atom scatterers the photon~\cite{Eichmann, Itano}.
The resulting entanglement fidelity is found to be
$F=1-4\eta^2(\bar{n}+1/2)$ in the Lamb-Dicke limit where $\eta^2
\bar{n} \ll 1$~\cite{cabrillo:1999}. Here, $\eta = \Delta k
\sqrt{\hbar/(2m\omega)}$ is the Lamb-Dicke parameter of each ion of
mass $m$, and $\bar{n}$ is the average number of thermal quanta of
motion in a trap of frequency $\omega$. This problem can be overcome
by collecting photons in the forward scattering direction ($\Delta k
=0$), or by confining the ions deeply within the Lamb-Dicke limit
where the recoil probability is small.

\subsection{Type II Heralded Entanglement and Gate Operation}
The interferometric stability requirement for type I entanglement is
a serious challenge for experimental implementation. Type II
entanglement bypasses this issue through interference of two photons, one from
each atom, where the interferometric phase $k\Delta x$ becomes common mode.
This makes type II entanglement more robust to noise, and it has been successfully
demonstrated in experiments
\cite{Moehring:Entangle:2007,Matsukevich:Bell:2008,Olmschenk:Tele:2009,Maunz:Gate:2009}.

Polarization, frequency, and time--bin qubits can all be used for
type II entanglement generation. Frequency and time--bin qubits can
also be used to implement a heralded quantum gate, where the output
state depends on the input state of the atomic qubits. We discuss
this gate operation, and show how it can be used for entanglement
generation.

Initially atoms $A$ and $B$ are independently prepared in arbitrary
quantum states. The total atom--atom state is described by
\begin{equation}
|\Psi_{aa}\rangle_i=\left(\alpha_A|\uparrow\rangle_A+\beta_A|\downarrow\rangle_A)\otimes(\alpha_B|\uparrow\rangle_B+\beta_B|\downarrow\rangle_B\right).
\label{eq:gateini}
\end{equation}
Suppose the entangled atom--photon pairs are generated, preserving the initial quantum information by using appropriate photonic qubits. If we assume equal optical path lengths from each atom to the beamsplitter for simplicity, the overall input state of the atom--photon pairs to the beamsplitter is
\begin{eqnarray}
|\Psi_{apap}\rangle
&=&\big(\alpha_A|\uparrow\rangle_{A}|P_\uparrow\rangle_{A}+\beta_A|\downarrow\rangle_{A}|P_\downarrow\rangle_{A}\big)
\otimes\big(\alpha_B|\uparrow\rangle_{B}|P_\uparrow\rangle_{B}+\beta_B|\downarrow\rangle_{B}|P_\downarrow\rangle_{B}\big)\\
&=&|\tilde{\phi}^{+}\rangle_{aa}|\phi^{+}\rangle_{pp}
+|\tilde{\phi}^{-}\rangle_{aa}|\phi^{-}\rangle_{pp}
+|\tilde{\psi}^{+}\rangle_{aa}|\psi^{+}\rangle_{pp}
+|\tilde{\psi}^{-}\rangle_{aa}|\psi^{-}\rangle_{pp},
\label{eq:gatemid}
\end{eqnarray}
where $|\phi^{\pm}\rangle_{pp}$ and $|\psi^{\pm}\rangle_{pp}$ are the
maximally entangled Bell states for photonic qubits and $|\tilde{\phi}^{\pm}\rangle_{aa}$ and $|\tilde{\psi}^{\pm}\rangle_{aa}$ are the associated atomic states given by
\begin{eqnarray}
|\phi^{\pm}\rangle_{pp}&=&\frac{1}{\sqrt{2}}\big(|P_\uparrow\rangle_{A}|P_\uparrow\rangle_{B}\pm|P_\downarrow\rangle_{A}|P_\downarrow\rangle_{B}\big)\,\\
|\psi^{\pm}\rangle_{pp}&=&\frac{1}{\sqrt{2}}\big(|P_\uparrow\rangle_{A}|P_\downarrow\rangle_{B}\pm|P_\downarrow\rangle_{A}|P_\uparrow\rangle_{B}\big)\,\\
|\tilde{\phi}^{\pm}\rangle_{aa}&=&\frac{1}{\sqrt{2}}\big(\alpha_A\alpha_B\mid\uparrow\rangle_A\mid\uparrow\rangle_B\pm\beta_A\beta_B\mid\downarrow\rangle_A\mid\downarrow\rangle_B\big)\,\\
|\tilde{\psi}^{\pm}\rangle_{aa}&=&\frac{1}{\sqrt{2}}\big(\alpha_A\beta_B\mid\uparrow\rangle_A\mid\downarrow\rangle_B\pm\beta_A\alpha_B\mid\downarrow\rangle_A\mid\uparrow\rangle_B\big).
\label{eq:bellstate}
\end{eqnarray}

As a consequence of the quantum interference at the beamsplitter,
the photons will emerge from different exit ports only if they are
in the antisymmetric state $|\psi^{-}\rangle_{pp}$
\cite{hong:1987,Shih:1988,braunstein:1995,mandel:1999}. Using
frequency qubits as an example, where
$|P_\uparrow\rangle=|1_r0_b\rangle$ and
$|P_\downarrow\rangle=|0_r1_b\rangle$, according to
Eq.~\ref{eq:onep}, the beamsplitter produces the output state
\begin{equation}
|\psi^{-}\rangle_{pp}=\frac{1}{\sqrt{2}}(|1_r0_b\rangle_{A}|0_r1_b\rangle_{B}-|0_r1_b\rangle_{A}|1_r0_b\rangle_{B}).
\label{eq:gatefinal}
\end{equation}
Consequently, when the two PMTs detect a coincident event,
the final atom--atom state $|\Psi_{aa}\rangle_f$ is projected into
\begin{equation}
|\Psi_{aa}\rangle_f=\frac{\alpha_A\beta_B|\uparrow\rangle_A|\downarrow\rangle_B-\beta_A\alpha_B|\downarrow\rangle_A|\uparrow\rangle_B}{\sqrt{|\alpha_A\beta_B|^2+|\beta_A\alpha_B|^2}}.
\label{eq:gatefinal2}
\end{equation}

The above protocol generates the final state $|\Psi_{aa}\rangle_f$ from the initial state $|\Psi_{aa}\rangle_i$ by a gate operation given by
\begin{equation}
\frac{1}{2}Z_A\,(I-Z_A\,Z_B)=\left(\begin{array}{cccc}
0 & 0 & 0 &0\\
0 & 1 & 0 &0\\
0 & 0 &-1 &0\\
0 & 0 & 0 &0\end{array} \right)\,\,,
\label{eq:heraldedgate}
\end{equation}
where $Z_{A(B)}$ is the single qubit Pauli-z gate for atom $A$
($B$). Notice that this gate is not a unitary operator since the
input states $\mid\uparrow\rangle_A|\uparrow\rangle_B$ and
$\mid\downarrow\rangle_A|\downarrow\rangle_B$ do not output any
heralded events, yielding a null result. This quantum gate operation
can be used to entangle two atomic qubits: for
example, if the initial state is set to
$\alpha_A=\alpha_B=\beta_A=\beta_B=1/\sqrt{2}$ in
Eq.~\ref{eq:gateini}, the gate operation generates a maximally
entangled Bell state.

There are some advantages to performing type II entanglement with
time--bin qubits. First, in contrast to frequency and polarization
qubits, photons in time--bin qubits have only one component mode of
polarization and frequency and are less sensitive to birefringence
and dispersion in the photonic channel. This feature also makes
time--bin qubits a good candidate for generating entangled
atom--photon pairs using a cavity setup as discussed in
Sec.~\ref{sec:cavity}. Second, the system can be projected into
either the $|\psi^{+}\rangle_{pp}$ or the $|\psi^{-}\rangle_{pp}$
state by using time--bin qubits, where
$|P_\uparrow\rangle=|0_{t_1}1_{t_2}\rangle$ and
$|P_\downarrow\rangle=|1_{t_1}0_{t_2}\rangle$. Note that this is a
consequence of PMTs being able to resolve the arrival time of
photons but not the frequency. From Eq.~\ref{eq:onep}, the output
states from the beamsplitter for $|\psi^{\pm}\rangle_{pp}$ are given
by
\begin{eqnarray}
|\psi^{-}\rangle_{pp}&=&\frac{1}{\sqrt{2}}\left(|1_{t_1}0_{t_2}\rangle_{A}|0_{t_1}1_{t_2}\rangle_{B}
                                -|0_{t_1}1_{t_2}\rangle_{A}|1_{t_1}0_{t_2}\rangle_{B}\right)\,,\\
|\psi^{+}\rangle_{pp}&=&\frac{1}{\sqrt{2}}\left(|0_{t_1}0_{t_2}\rangle_{A}|1_{t_1}1_{t_2}\rangle_{B}
                                -|1_{t_1}1_{t_2}\rangle_{A}|0_{t_1}0_{t_2}\rangle_{B}\right).
\label{eq:time-bin}
\end{eqnarray}
This explicitly shows that if only one scattered photon
is detected at each time $t_1$ and $t_2$ by different PMTs, the
photon state is projected into the $|\psi^{-}\rangle_{pp}$ state,
which projects the atom--atom state into the entangled state
$|\tilde{\psi}^{-}\rangle_{aa}$. Instead, if the events are detected
by the same PMT, the photon state is projected into the
$|\psi^{+}\rangle_{pp}$ state, yielding the atom--atom state
$|\tilde{\psi}^{+}\rangle_{aa}$.

Since type II entanglement schemes require the presence of two
photons, the success probability is
$P_{II}=p_{B}(P_{ap}\,\eta_{det})^2$, where $p_B$ is the probability
of detecting a Bell state of the two photons (in the examples above,
$p_B=1/4$ for the frequency qubit and $p_B=1/2$ for the time-bin
qubit). The success probability for type II entanglement is
quadratic in $P_{ap}$ and is thus typically much smaller than that
of type I for $P_{ap} \ll 1$. However, type II entanglement has the
advantages of no fundamental fidelity limit and far less sensitivity
to experimental noise and interferometric instability.

If the path length difference between the two photonic channels is
offset by an amount $\Delta x$, a phase factor $e^{i\,\Delta
\omega\,\Delta x/c}$ emerges between the two components of
Eq.~\ref{eq:gatefinal2}. For frequency photonic qubits, $\Delta
\omega$ is the difference frequency between the two photon
frequencies, and for time-bin qubits this is given by the frequency
difference between the atomic qubit levels. Since $\Delta \omega/c
\ll k$, type II entanglement and gate operation are typically much
more robust than type I schemes.

\section{Quantum Information Protocols Based on Multi-ion Emission\label{sec:multiion}}

Photon emission from multiple ions can be a useful technique to
scale up the type I entanglement protocol to create large
multi-partite entangled states. In this section we focus on the type
I scheme to entangle many atoms through the collective emission of
one photon. This entanglement protocol can be used to generate
multi-partite entangled W-states~\cite{cabrillo:1999, kielpinski}.
W-states have been shown to provide a means to secure quantum
communication~\cite{Joo}, and have also been proven to be the only
state capable of solving certain problems on quantum
networks~\cite{D'Hondt}.  Besides these specific applications, the
W-state may be of interest for investigating emergent
phenomena in large multi-partite entangled states\cite{Davies}.

$N$-particle W-states of the form $|\tilde{W}_N\rangle =
\frac{1}{\sqrt{N}}(e^{i\phi_1}|00....01\rangle+e^{i\phi_2}|00....10\rangle+....+e^{i\phi_N}|10....00\rangle)$
can be created by weakly exciting $N$ atoms such that the probability to scatter
more than one photon is vanishingly small~\cite{cabrillo:1999}.  The notation
$|\tilde{W}_N\rangle$ refers to generalized W-states, in which the
phase factors present in the quantum state are arbitrary but fixed. The
specific state where all the phase factors are equal is simply
referred to as the W-state $|W_N\rangle$.
Implementing such a type I process with remotely-located ions will come at the cost of
maintaining optical interferometric stability, so we consider the heralding of
W-states by using atoms confined in a single trap.
The generation of a N-atom W-state can be verified by measuring the entanglement
witness given by
$\hat{\mathcal{W}}=\frac{N-1}{N}\hat{\mathbf{I}}-|W_N\rangle\langle
W_N|$, the negative expectation value of which signals multi-partite
entanglement \cite{Bourennane}. By exploiting the symmetry of the W-state with respect
to an exchange of any two qubits, it was shown that this witness
operator can be measured by $2N-1$ measurements without individual
addressing~\cite{Guhne}. Because the phase factors of the W state are determined by the
relative optical path lengths from each ion to the detector, the angular
distribution of the scattered photon forms interference fringes.
In order to use this witness to characterize the entanglement, the
photon detector must be able to resolve these fringes.

We develop a method of calculating the fidelity of the W-state
by drawing a parallel between the storage of coherence in the
atoms through an inelastic scattering event and the coherence in the
light field through an elastic scattering event. In the experiment
of Ref.~\cite{Eichmann}, where two ions in a single trap were weakly
excited, the position of scattered photons was shown to exhibit
interference in elastic scattering events but no interference in
inelastic scattering events. In the case of elastic scattering, no
information about which ion scattered the photon will exist,
allowing the different optical paths to
interfere~\cite{Eichmann,Scully}. This process yields an
interference pattern with regions of high intensity, where the
photon phases $e^{-i\vec{k}\cdot \vec{R}}$ add constructively. In
the case of inelastic scattering, the photon phase
$e^{-i\vec{k}\cdot \vec{R}}$ gets imprinted on the ions upon
detection of a photon.  The ions will also pick up a dynamical phase
$e^{-i\omega_A \Delta t}$, where $\Delta t$ is the time it takes for
the photon to reach the detector, and $\omega_A$ is the frequency
difference between $|0\rangle$ and $|1\rangle$. If the time that
photons take to traverse the full length of the ion crystal is short
compared to ${\omega_A}^{-1}$, this dynamical factor can be ignored.
For the ${}^{171}\mathrm{Yb}^+$ Zeeman splitting with a magnetic
field on the order of a few Gauss, this approximation is valid for
ion crystals much smaller than one meter.  When this approximation
is valid, the regions of high intensity in the \textit{elastic} case
are the same points at which the detection of a photon in the
\textit{inelastic} case signals the creation of a W-state with all
the phases being equal. Therefore, in order to identify scattering
regions that herald a high fidelity W-state, it should suffice to
calculate the elastic scattering cross-section for $N$ ions in a
trap and identify the high intensity region.

In order to find the points of high intensity in the radiation
pattern, we generalize the derivation of the scattering
cross-section for two ions in a single trap~\cite{Itano} to $N$
ions.  Starting with the differential scattering cross-section as given by the
electric dipole Hamiltonian in second-order perturbation theory
\begin{equation}
\frac{d\sigma}{d\theta} =\sum_f\left|\sum_{p,j}
\frac{\langle\Psi_f|(\textbf{D}_p\cdot\hat{\epsilon}_{out})\,e^{-i\vec{k}_{out}\cdot\textbf{R}_p}\,|\Psi_j\rangle\langle\Psi_j|\,(\textbf{D}_p\cdot\hat{\epsilon}_{in})\,e^{i\vec{k}_{in}\cdot\textbf{R}_p}\,|\Psi_i\rangle}
{\omega_0-\omega_{in}+(E_j - E_i)/\hbar - i\Gamma/2} \right|^2\,,
\label{eq:perturb}
\end{equation}
where $\textbf{D}_p$ and $\textbf{R}_p$ are the dipole and position
operators for the $p^{\,\mathrm{th}}$ ion, $\hat{\epsilon}$ is a
polarization vector, $\vec{k}_{in}$ and $\vec{k}_{out}$ are incoming
and outgoing wavevectors and the indices $i$ and $f$ represent the initial and final
state of the ions. It is important to note that this expression only
applies to elastic scattering events, since in this case the
probability amplitudes for the photon scattering off different ions
are added together because these processes are indistinguishable.
The quantum states in the perturbation expansion are taken to be
eigenvectors of the unperturbed Hamiltonian of $N$ ions in a
harmonic trapping potential and are therefore product states of the
ions' internal degrees of freedom and the motional state of the
system.

We now examine the special case of the ion crystal axis lying in the
plane defined by the incoming and outgoing wavevectors and the
quantization axis being perpendicular to that plane.  In this case,
the dipole operator only contributes an overall scaling factor to
the scattering cross-section and is therefore ignored in this
calculation.  As explained in~\cite{Itano}, the denominators in
Eq.~\ref{eq:perturb} are nearly constant for the
${}^{171}\mathrm{Yb}^+$ ion cooled near the Doppler limit on the
$370$~nm line as a consequence of the recoil frequency, $8.5$~kHz,
being small compared to the linewidth.  By considering the
denominators to be nearly constant, they can be factored out of the
sum allowing us to do the sums over the $j$ and $f$ indices via
resolutions of the identity operator, giving
\begin{equation}
\frac{d\sigma}{d\theta}=\langle\{n_{HO}\}_i|\sum_{p,p'}e^{-i \left(\vec{k}_{out}-\vec{k}_{in}\right)\cdot\left(\textbf{R}_{p}-\textbf{R}_{p'}\right)}|\{n_{HO}\}_i\rangle,
\label{eq:interference_pattern}
\end{equation}
where $\{n_{HO}\}_i$ denotes the initial harmonic oscillator quantum
numbers of the ions. Eq.~\ref{eq:interference_pattern} might be seen
as a classical interference pattern from a diffraction grating
comprised of slits that oscillate around fixed points and recoil
upon deflection of light quanta.  After taking the expectation
value, the final scattering cross-section is given by
\begin{equation}
\frac{d\sigma}{d\theta}=\sum_{p,p'}e^{i\eta_\lambda
(U_p-U_{p'})\Delta\hat{k}\cdot\hat{x}} \prod_{m=1}^{N}
e^{-[(A_{p,m}-A_{p',m})\eta_a^m\Delta\hat{k}\cdot\hat{x}]^2(\bar{n}^m_a
+ 1/2)
+[(T_{p,m}-T_{p',m})\eta_t^m\Delta\hat{k}\cdot\hat{y}]^2(\bar{n}^m_t+1/2)}.
\label{eq:scatteramp}
\end{equation}
In the above equation, $\bar{n^m_a}(\bar{n^m_t})$ are the average
number of axial (transverse) thermal quanta in mode $m$ and
$\eta_\lambda\equiv |\vec{k}|d$. $U$ is a vector containing the
equilibrium positions of the ions in units of length
$d=\left[e^2/m(\omega^1_a)^2\right]^{1/3}$. The Lamb-Dicke parameter
for the $m^{\mathrm{th}}$ mode in the axial (transverse) direction
is denoted
$\eta_{a(t)}^m=|\vec{k}|(\hbar/(2m\omega_{a(t)}^m))^{1/2}$. The
difference between incoming and outgoing wavevectors is
$\Delta\hat{k}=\hat{k}_{out}-\hat{k}_{in}$, and $A_{p,m},\,T_{p,m}$
are elements of transformation matrices from position coordinates to
axial and transverse normal coordinates, respectively.

The fidelity of the entangled state with arbitrary phase of two ions
in separate isotropic harmonic traps (frequency $\nu$ and average
thermal index $\bar{n}$) as derived in Ref.~\cite{cabrillo:1999} is
given by
\begin{equation}
F(\theta)=\int_0^{\infty}d\tau e^{-\tau}e^{-4\eta^2(\bar{n} + 1/2)
[1-\mathrm{cos}(\chi)\mathrm{cos}(\frac{\nu\tau}{\Gamma})]},
\label{eq:Cabrillo_fidelity}
\end{equation}
with $\eta=k\sqrt{\hbar/(2m\nu)}$ and $\chi$ being the angle between
the excitation beam and the emission direction. In the limit of weak
confinement, $\nu\ll\Gamma$, and because the integrand decays
exponentially in $\tau$, we can approximate
$\mathrm{cos}\left(\frac{\nu\tau}{\Gamma}\right)\approx 1$, allowing
us to carry out the integration and arrive at
\begin{equation}
F(\theta) \approx e^{-4\eta^2(\bar{n} + 1/2)
[1-\mathrm{cos}\left(\chi\right)]}= e^{-8\eta^2(\bar{n} +
1/2)[(\Delta\hat{k}\cdot\hat{x})^2+(\Delta\hat{k}\cdot\hat{y})^2]}\,.
\end{equation}
This shows the relationship between Eq.~\ref{eq:Cabrillo_fidelity}
and the contrast of the fringes in Eq.~\ref{eq:scatteramp}. This
suggests that the fringe contrast of Eq.~\ref{eq:scatteramp} for
$N=2$ might be interpreted as the fidelity of the entangled state
with arbitrary phase when two ions are in the same trap.  Moreover,
because the fringe peaks correspond to points of common phase, the
full expression (Eq.~\ref{eq:scatteramp}) for $N=2$ might
be interpreted as the fidelity of the state where the relative phase
is equal to zero. We contend that Eq.~\ref{eq:scatteramp} should be
a valid prediction of the fidelity of an $N$ qubit W-state in the
weak confinement regime, with $\eta_\lambda\gg1$ ensuring no
ion--ion photon exchange.

\begin{figure}[tb]
\begin{center}
\includegraphics[width=2.5in]{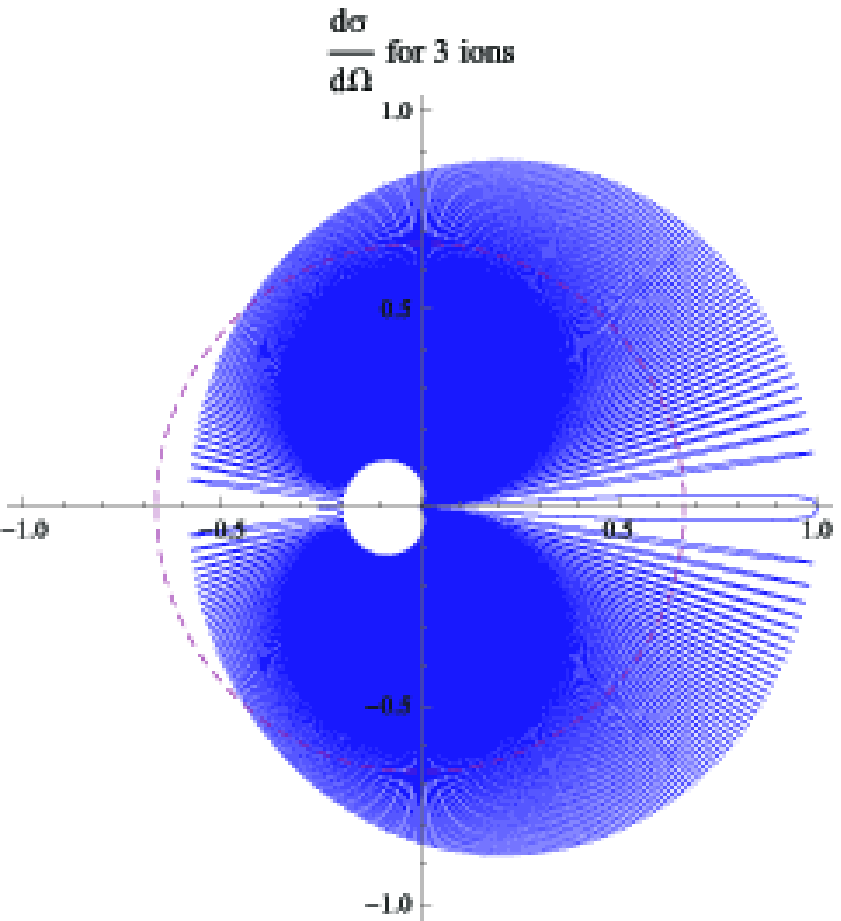}
\includegraphics[width=2.5in]{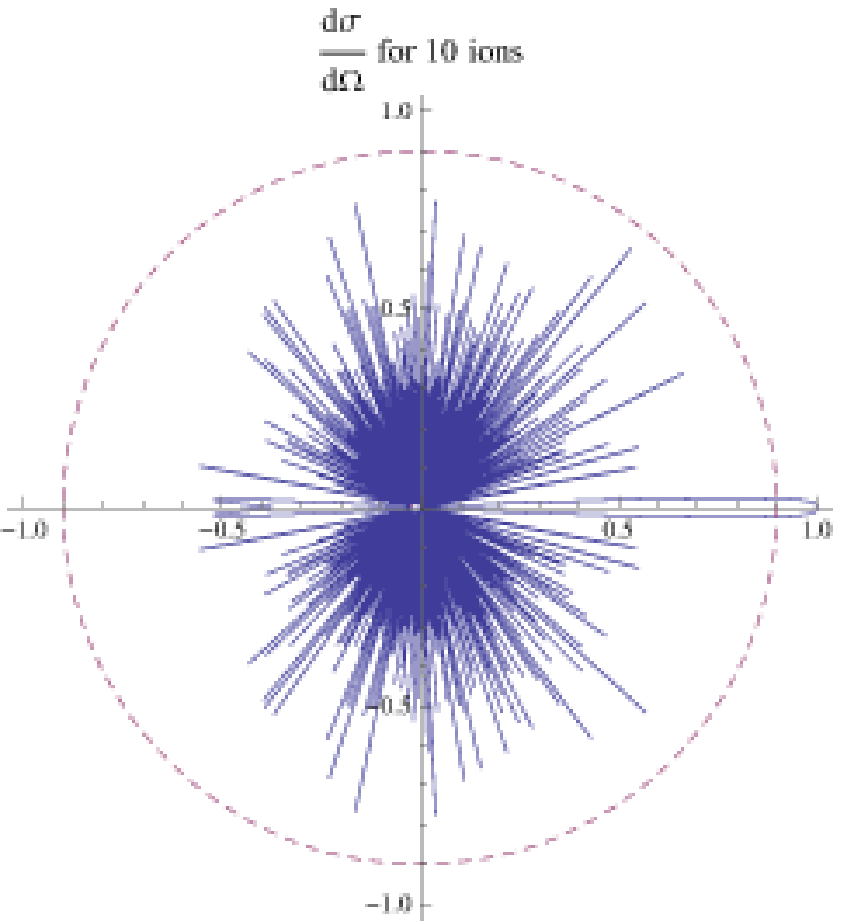}
\caption{(a) The differential scattering cross section, in the plane
defined by the ion crystal axis and the excitation vector, is shown
in the polar plot with the excitation beam coming in along the ion
crystal axis at $\theta=0$.  The dashed line represents the
normalized intensity (2/3) required to expect a negative expectation
value of the entanglement witness W-state entanglement witness.  For
the example shown here, realistic trap parameters were chosen and
Doppler-limited cooling was assumed giving
$\eta_\lambda=600,\sqrt{\frac{\hbar k^2\Gamma}{2m\omega_x^2}}=1$,
and $\frac{\omega_x}{\omega_y}=10$. (b) The scattering cross section
for ten ions in a harmonic trap is plotted together with the
required normalized intensity (0.9, dashed circle) for a negative
expectation value of the witness operator.}\label{fig:Ion Radiation}
\end{center}
\end{figure}

In Fig.~\ref{fig:Ion Radiation}(a) the emission pattern for three
ions is plotted, clearly showing the degrading effect of recoil due
to a large scattering angle. As a consequence of the ions not being
evenly spaced, the scattering profile from ten ions in
Fig.~\ref{fig:Ion Radiation}(b) shows that the only points where
radiation adds up in phase is in the forward scattering direction.
This implies that in an inelastic scattering event, the only points
where light detection will yield a W-state with all the terms having
the same phase is in the forward scattering direction.  The angular
size of this spot, $\delta\theta$, in the case where the excitation
pulse is along the crystal axis can be estimated by first
normalizing Eq.~\ref{eq:scatteramp} by dividing by $N^2$, ignoring
the Debye-Waller factors and making the small scattering angle
approximation, giving
\begin{equation}
\frac{d\sigma}{d\theta}\approx1-\frac{\delta\theta^4 \eta_\lambda^2}{4
N^2}\sum_{p>p'}{(U_{p}-U_{p'})^2}.
\label{eq:forward_scatter}
\end{equation}
The sum in Eq.~\ref{eq:forward_scatter} can be approximated by
numerically solving for the equilibrium positions for different
numbers of ions in a harmonic trap, which yields
$\sum_{p>p'}{(U_{p}-U_{p'})^2}\approx0.45N^{2.87}$.  We define the
spot size to be the region where the intensity is at least $f$ times
the maximum and find the angular size of the spot to be
approximately given by
\begin{equation}
2\,\delta\theta\approx2\frac{1.7(1-f)^{1/4}}{\eta_\lambda^{1/2}}N^{-0.21}.
\label{eq:spot_size}
\end{equation}
Remembering that the entanglement witness demands that the fidelity
of the W-state be bounded by $F\geq\frac{N-1}{N}$, we set
$f=\frac{N-1}{N}$. With the interpretation that the elastic
scattering cross-section represents the fidelity of the W-state for
an inelastic scattering event, the fraction of photons scattered
into the plane of interest that will yield an $N$ particle W-state
is solved
\begin{equation}
\frac{2\delta\theta}{2\pi}\approx\frac{0.55}{\eta_{\lambda}^{1/2}}N^{-0.46}.
\label{eq:available_fraction}
\end{equation}
Eq.~\ref{eq:available_fraction} allows the estimation of an upper
bound on the angle subtended by the detector being used to signal
the creation of a multi-partite entangled state.  This scaling law
shows that the efficiency with which one can create multi-partite
entanglement decreases rather slowly with the number of entangled
ions.


\section{Techniques for Enhancing Light Collection \label{sec:MethodsLC}}

Collecting spontaneously emitted photons from trapped atomic ions
traditionally involves the use of refractive optics in free space. A
common setup uses a high numerical aperture objective to collect
light for entanglement protocols as well as state detection.  For
the entanglement protocols discussed in Sec.~\ref{sec:protocol},
mode matching of the photons is critical, so light is typically coupled into
single-mode optical fibers. For this type
of free-space light collection, there are two critical difficulties
that limit the collection efficiency. First, the solid angle
subtended by the collecting lens, $\Delta \Omega/(4 \pi)$, is
usually on the order of $10^{-2}$\cite{Olmschenk:Tele:2009}. This
small collection angle contributes a factor of order $10^{-4}$ to
the total success probability for type II heralded entanglement
schemes. Second, the mode overlap of the optical dipole mode to the
fiber mode substantially decreases the amount of light transmitted
by the optical system.

\begin{figure}[tb]
\includegraphics[width=5.5 in]{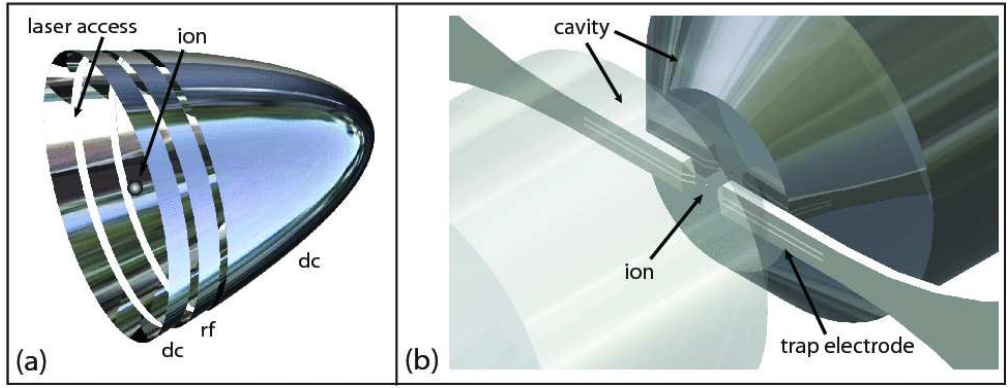}
\caption{Schemes to enhanced light collection from trapped ions.(a)
A parabolic geometry for a novel ion trap, where the RF and DC trap
electrodes constitute the reflective surface. The paraboloid is
segmented such that the RF node is positioned at the focus of the
mirror. (b) A 3D RF quadrupole trap positioned between two cavity
mirrors. The electrodes of the trap are small enough to fit between
the mirrors of the optical cavity. Each substrate has a narrow RF
center electrode, while two outer electrodes provide a close ground
for both the RF power and charge that may accumulate on the
dielectric surfaces. The electrodes can be independently mounted on
movable stages controlled via vacuum feedthroughs to precisely
position the ion \emph{in situ}.} \label{fig:Iontrap}
\end{figure}

In recent years many proposals for enhancing atom--photon coupling
have emerged
~\cite{mckeever:2004,Lindlein,Rempe:server:2007,Kurtsiefer:freespace:2008,Blatt:inter:2009,blinov:0901.4742}.
In Fig.~\ref{fig:Iontrap} we show two possibilities of employing
either reflective optics or an optical cavity to increase light
collection efficiency. One option is to use a single parabolic
mirror to collect a large fraction of the scattered light from the
atom~\cite{Lindlein}. Another method is to place the atom inside a
high finesse optical cavity and utilize the Purcell effect to
extract photons. For this application, the strong coupling regime is
not necessary. Instead, the perturbative regime, or ``bad cavity''
limit (Sec.~\ref{sec:cavity}), will suffice, as long as the coherent
coupling rate is larger than the dipole decay rate.

Ion traps suitable for such experiments would necessarily need to
have an optically open geometry, such as a surface
trap~\cite{Chiaverini:2005,Seidelin:2006} or a needle
trap~\cite{deslauriers:2006,wineland:0810.2647}, to minimize the
occlusion of light reflected from a mirror or the optical mode of
the cavity. The radio frequency (RF) node of the ion trap must be
precisely placed at the focus of the reflective mirror or along the
cavity axis.  The ion trap could be monolithically created, ensuring
precise alignment of the RF node with the optical field, or it could
be aligned \emph{in situ}. For dielectric optics, the characteristic
ion--electrode distance must be much smaller than the ion--optic
spacing in order to provide adequate shielding of the ion from
potential charge accumulation on the optic. Fig.~\ref{fig:Iontrap}
is a conceptual view of an ion confined at the the focus of a parabolic reflector
trap (Fig.~\ref{fig:Iontrap}a) or in an optical cavity
(Fig.~\ref{fig:Iontrap}b).

\subsection{Reflective Optics}

One way of collecting more light emitted by a single trapped ion is
to place the minimum of the trapping potential at the focus of a
parabolic mirror. In this section we analyze the fiber-coupling
efficiency of the dipole radiation from an ion at the focus of such
a trap, with its quantization axis along the line of symmetry.

Prior to reflection, the optical fields of emitted photons
associated with the three possible transitions are given by
\begin{equation}
\vec{E}_{l=1,m=0}=
\frac{ie^{ikr}}{r}\sqrt{\frac{3}{16\pi}}\mathrm{sin}\theta\hat{\theta},\,\,\,\,\,
\vec{E}_{l=1,m=\pm 1}= \frac{ie^{ikr}}{r}e^{\pm
i\phi}\sqrt{\frac{3}{16\pi}}(\pm \mathrm{cos} \theta\hat{\theta}+
i\hat{\phi})\,.
\end{equation}
Once the light has reflected and formed a collimated beam, the
wavefronts are flat, meaning the exponential factor will be a
constant and can be ignored.

\begin{figure}[tb]
\begin{center}
\includegraphics[width=2.5in]{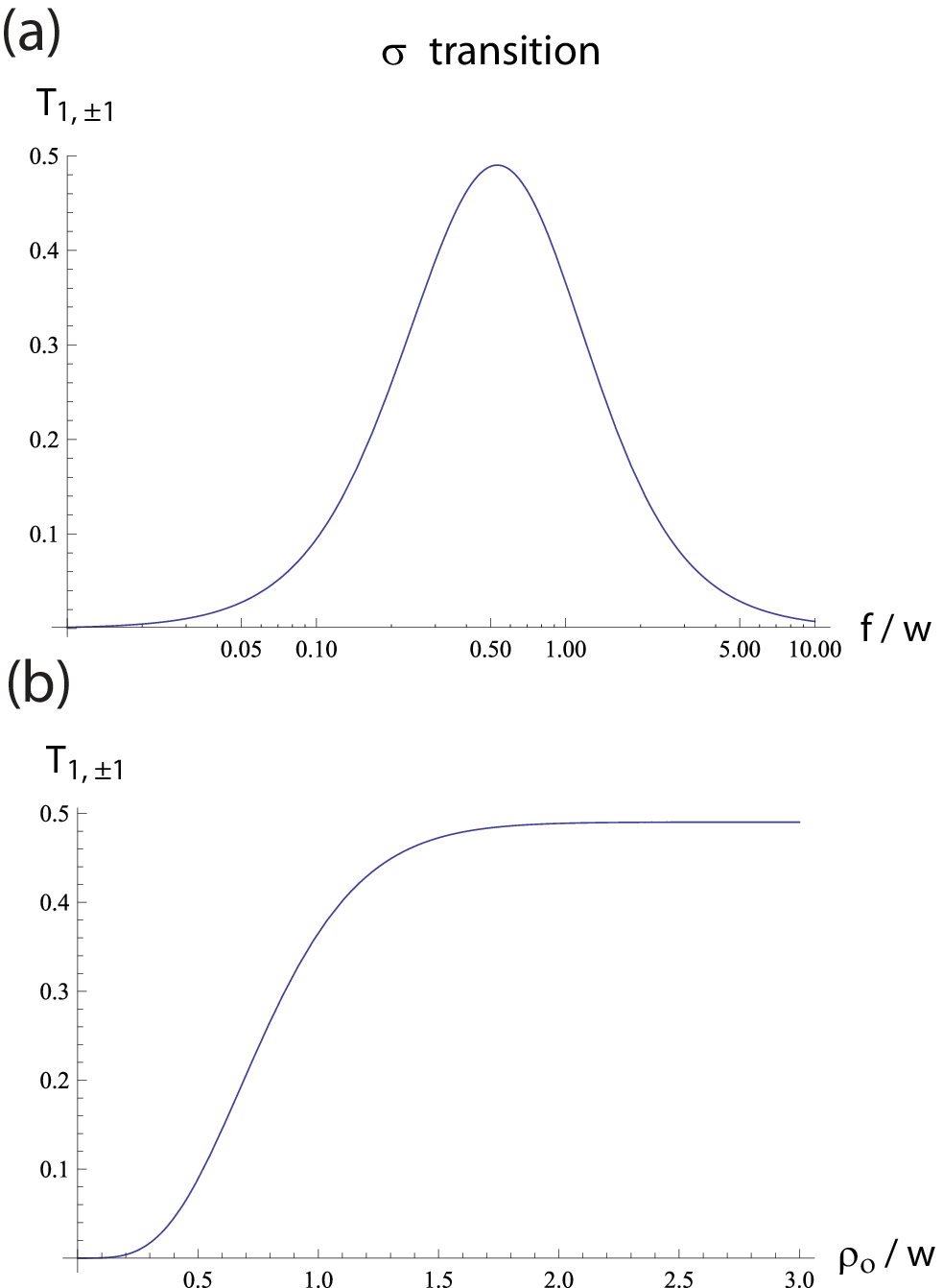}
\includegraphics[width=2.5in]{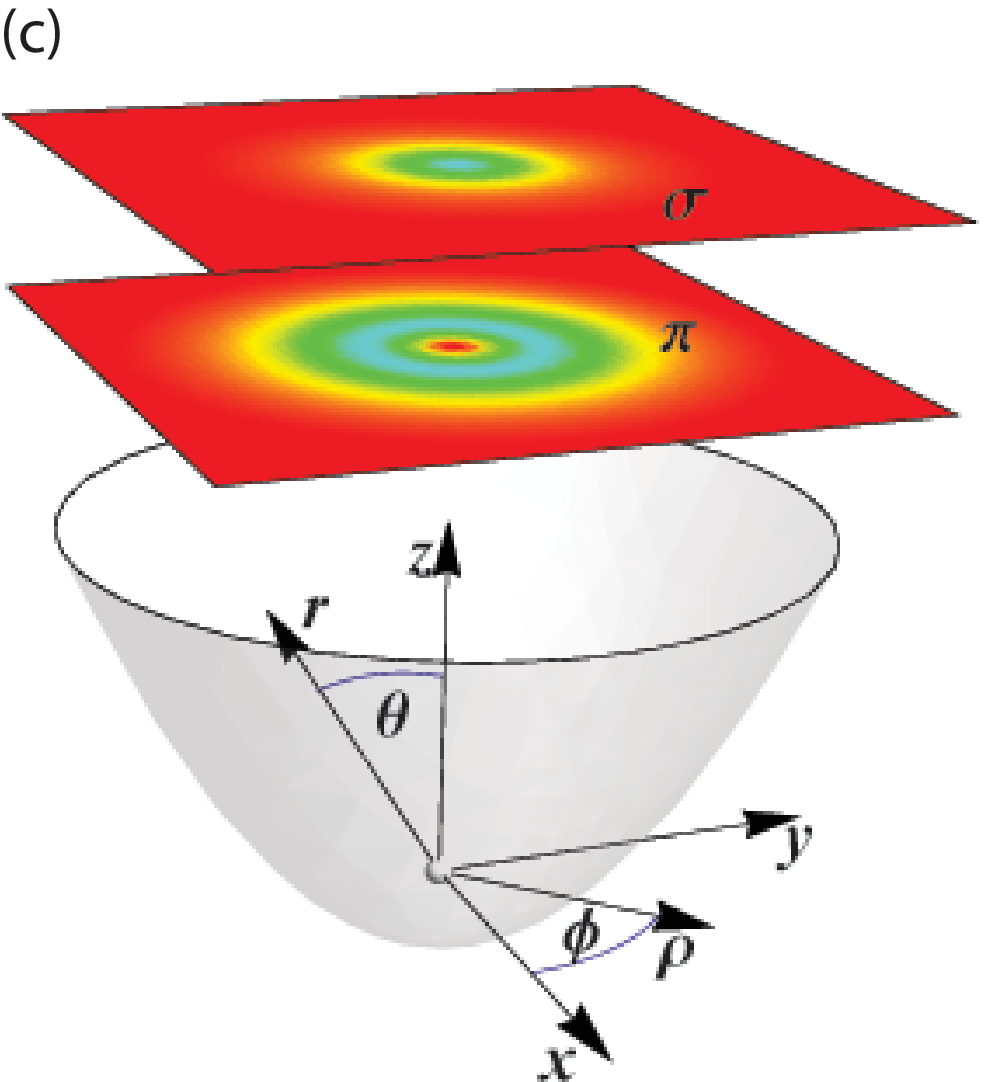}
\caption{Light collection using an infinite parabolic mirror to
reflect the emitted photons. (a) The probability of collecting a
photon from a $\sigma$ transition in a single-mode fiber as a
function of $f/w$. (b) The probability of collecting a photon in a
single-mode fiber as a function of the maximum radius of the
paraboloid (with an optimal focus) in units of $w$. (c) The
intensity distributions of $\pi$ and $\sigma$ light after reflected
off a parabolic mirror.  It clearly shows that the $\pi$ light forms
a ``doughnut'' mode while the $\sigma$ light more closely resembles
distribution of a Gaussian beam.} \label{fig:reflection_coupling}
\end{center}
\end{figure}

Using polar coordinates shown in Fig.~\ref{fig:reflection_coupling}, the
paraboloid is defined by $z(\rho,f)=\frac{\rho^2}{4f}-f$, and the
distance from the ion to the reflecting surface is then given by
$(\rho^2+4f^2)/4f$. After reflection, light that was polarized in
the $\hat{\theta}$ direction will have polarization in the
$-\hat{\rho}$  direction, while the $\hat{\phi}$ polarization is unchanged.
The intensity profile that was only a function of
$\theta$ will be mapped to an intensity profile that only depends on
$\rho$. The mapping can be found by solving for the distance from
the symmetry axis after reflection as a function of the angle of
emission, giving $\theta=\mathrm{tan}^{-1}[(4f\rho/(\rho^2-4f^2)]$.

As shown in Eq.~\ref{eq:m0mode}, the $m=0$ transition will
produce a donut mode with radial polarization given by
\begin{equation}
\vec{E}_{l=1,m=0}\rightarrow
-\frac{i4f}{\rho^2+4f^2}\sqrt{\frac{3}{16\pi}}\frac{4f\rho}{\rho^2+4f^2}\hat{\rho}\,\,,
\label{eq:m0mode}
\end{equation}
which is consistent with the optimal intensity profile for driving a
$\pi$ transition derived in \cite{Lindlein}. The amount of light from this mode that will couple into a single mode fiber is given by the mode overlap
\begin{equation}
T_{l,m}=\frac{|\int_0^{2\pi}\int_0^{\rho_o} d\phi\, d\rho\,\rho\,
\vec{E}_{l,m}\cdot\vec{G}|^2}{\int_0^{2\pi}\int_0^{\infty} d\phi\,
d\rho\, \rho\,
\vec{E}_{l,m}\cdot\vec{E}_{l,m}\int_0^{2\pi}\int_0^{\infty} d\phi\,
d\rho\, \rho \,\vec{G}\cdot\vec{G}}\,,
\end{equation}
where the quantity $\rho_o$ is the maximum radius of the mirror
being used and $\vec{G}=e^{-(\rho/w)^2} (\alpha \hat{x}+\beta
\hat{y})$ is the Gaussian mode of the fiber where
$|\alpha|^2+|\beta|^2=1$ .  Assuming the intensity and polarization
of the Gaussian mode only depend on $\rho$, the only dependence of
the integrand in the numerator on the angle $\phi$ is in the unit
vector $\hat{\rho}$, causing the integral to vanish.  Therefore,
without additional optics, there will be no transmission of $\pi$
transition light into a fiber aligned with the axis of symmetry.

After reflection, the field for a $m=\pm 1$ transition is given by
\begin{equation}
\vec{E}_{l=1,m=\pm 1}\rightarrow \pm\frac{i4f}{\rho^2+4f^2}e^{\pm
i\phi}\sqrt{\frac{3}{16\pi}}\left(-\frac{\rho^2-4f^2}{\rho^2+4f^2}\hat{\rho}\pm
i\hat{\phi}\right). \label{sigma:transition}
\end{equation}
The collimated beam has circular polarization in the center, becomes
increasingly elliptical going away from the center until it is
azimuthally polarized at a distance of $\rho=2f$, then becomes
elliptical again approaching a polarization orthogonal to the center
of the beam. The overlap integral is calculated using Eq.~\ref{sigma:transition} to give the following coupling efficiency
\begin{equation}
P_\sigma=T_{1,\pm
1}=\frac{3}{2}\left(\frac{2f}{w}\right)^6|\alpha\pm
i\beta|^2e^{2(2f/w)^2}\left|\Gamma\left(-1,\frac{4f^2}{w^2}\right)-\Gamma\left(-1,\frac{{\rho_0}^2+4f^2}{w^2}\right)\right|^2,
\end{equation}
where $\Gamma(a,b)$ is the incomplete Gamma function.
This expression shows that the light from the $\sigma^+$ transition
coupling to the fiber has left-handed polarization, and the light
generated from the $\sigma^-$ transition coupling to the fiber has
right-handed polarization. If we take the limit where the paraboloid
has infinite extent, we can numerically solve for the focus that
gives maximum coupling efficiency.
Fig.~\ref{fig:reflection_coupling} shows the results of this
calculation, which predicts a maximum coupling efficiency near
$50\%$.

This setup may also be used for the logic gate described earlier
using frequency qubits. Because the $\pi$ transition light does not
couple into the fiber without additional optics, we consider
$\sigma$ transitions.  In order to preserve any initial coherence
set up in the ion after excitation and emission, we can use the
$\sigma^+\pm\sigma^-$ transitions in the ion.  The fields resulting
from these transitions are superpositions of
Eq.~\ref{sigma:transition}. Because the fiber maps the $\sigma$
transition photons to circularly polarized photons while preserving
orthogonality, a $\sigma^++\sigma^-$ photon captured by the fiber
will have a linear polarization orthogonal to a $\sigma^+-\sigma^-$
photon captured by the fiber.  With this setup, we can still use the
${}^{171}\mathrm{Yb}^{+}$ clock states in the ground state manifold
to hold quantum information, but now we use $\sigma$ light to
couple to the $m_F=\pm1$ Zeeman states in the $|^2
P_{1/2},F=1\rangle$ manifold.  Here, the two states
$|\pm\rangle = \mid\uparrow\rangle \pm \mid\downarrow\rangle$
are coupled to the excited state with
$\sigma^+ \pm \sigma^-$ polarization, respectively. This
implies that after excitation and decay the internal state of the
ion will be correlated with both the frequency and polarization of
the emitted photon.
In order to use the parabolic mirror for the
gate, the polarization information should be erased to allow for the
interference on the beam splitter to occur. The distinguishability
between the $\sigma^++\sigma^-$ and $\sigma^+-\sigma^-$ transitions
can be erased with a linear polarizer oriented to block
$\hat{x}+\hat{y}$.  Assuming perfect coupling into the fiber, the
probability to see a coincidence is now given by $\approx
p_{B}(\eta\frac{1}{2}\frac{1}{2}\frac{1}{2})^2$, where the three
factors of $1/2$ come from maximum coupling efficiency,
Clebsch-Gordan coefficients, and the polarization erasure.


\subsection{Optical Cavity\label{sec:cavity}}

Another method to increase light collection is to surround the ion
with an optical
cavity~\cite{mundt:2002,Matthias:ioncavity:2004,Blatt:cavity:2009}. The field
mode modifications imposed by the mirrors change the spontaneous
emission of the ion such that it preferentially emits into the
cavity mode, depending on the coupling parameters. This phenomenon
is the well-known Purcell effect~\cite{berman}.

The dominant ion--cavity coupling parameters are the coherent
atom--field coupling, $g$, the free space ion spontaneous emission
rate, $\Gamma$, and the cavity decay rate, $\kappa$.  These
parameters define the cooperativity, $C \equiv g^2/ \kappa\Gamma$,
which is a measure of the atom-cavity coupling
with respect to the dissipative processes. The atomic
spontaneous emission rate is enhanced by a factor proportional to
$1+2C$. If we assume that the photon collection efficiency equals
the probability of a spontaneously emitted photon emerging from the
outcoupling mirror of the cavity, over experimental time scales, the
collection efficiency is given by
\begin{equation}
p_{c} =
\frac{T_{l}}{\mathcal{L}}\left(\frac{2\kappa}{2\kappa + \Gamma}\right)\left(\frac{2C}{1+2C}\right).
\end{equation}
The first factor, $T_{l}/\mathcal{L}$ is the fraction of light that
is transmitted by the outcoupling mirror, $T_{l}$, to the total
losses of the cavity $\mathcal{L}$. The second factor relates the
rate photons leave the cavity to the rate they leave the ion--cavity
system. The third factor is the fraction of light scattered into the
cavity mode. The ``bad cavity'' regime can be intuitively envisioned
as providing enough atom--cavity coupling to transfer the excitation
from the atom into an entangled atom--photon pair.  This photon then
leaves the cavity faster than the time it takes to be reabsorbed by
the atom and emitted into unwanted side modes.

An optical cavity can be used for all the protocols discussed in
Sec.~\ref{sec:protocol}.  Number qubits can be realized with the
coherent transfer of an excitation from the atom to a single mode of
the cavity~\cite{kuhn:2002,mckeever:2004,Blatt:cavity:2009}. Polarization
qubits can be realized with a cavity coupling to Zeeman levels of atoms lying
inside the linewidth of the cavity~\cite{Rempe:atomphoton:2007}.
Frequency qubits would need both frequency modes resonant with the
cavity.  A difficulty with realizing frequency qubits is the
dependence of the coherent coupling strength, $g$ on the mode
volume. In order to ensure that $\omega_{b}-\omega_{r}$ is an
integral multiple of the free spectral range it may be necessary
that the cavity length be large. However, a longer cavity tends to
have a larger mode volume, leading to a smaller coupling strength
$g$. A method to combat this is a near-concentric geometry. In this
case, the centers of curvature of the two mirrors nearly coincide.
This leads to an optical mode with an extremely tight focus,
allowing a relatively strong coupling strength $g$. Time-bin qubits
can use the frequency selectivity of the cavity to ensure that one
of the transitions is off-resonant~\cite{barrett:2005}. However, to
preserve the coherence, one must be careful not to excite both
levels.  That is, the excitation pulse must have a bandwidth much
smaller than the qubit spacing, yet the bandwidth must be greater
than the excited state linewidth.


\section{Outlook and Conclusions\label{sec:outlook}}

The quantum network generated by photon-mediated entanglement can be
used for quantum communication and distributed quantum computation.
We list some possible applications for such an atom--photon network.

\textbf{Loophole-free Bell Inequality Test.} Loophole-free Bell
inequality tests are of continuing interest for testing fundamental
aspects of quantum mechanics\cite{PhysRevLett.23.880}. Bell
inequality violation experiments are subject to two primary
loopholes: the detection loophole, and the locality
loophole~\cite{aspect:1982b,rowe:2001}. While trapped ions typically
close the detection loophole due to nearly perfect state detection,
they have yet to close the locality loophole. On the other hand,
photonic qubits enable the separation required to close the
locality loophole, yet cannot be detected efficiently enough to
close the detection loophole. Combining the advantages of both
trapped ion and photons, the photon-mediated entanglement schemes
discussed above can generate an entangled ion pair that could close
both loopholes~\cite{simon:2003}. To close the locality loophole,
the remote ions must to be space-like separated; that is, the ions
must be separated greater than the distance
light travels in the time it takes to perform state detection.
For example, if the state detection time
were $10~\hbox{\textmu}$s, then the ions would need to be separated
by $3~\mathrm{km}$.  Since the starting point for a Bell inequality
measurement is the heralding of the remote entanglement, the success
probability does not play a role in
determining the separation of the two ions. However, the
improvements to the photon collection efficiency, $p_{c}$, can
shorten the state detection time, thus reducing the space separation
between two ions.

\textbf{Remote Deterministic Quantum Gates and Quantum Repeaters.}
The photon-mediated entanglement schemes can also be used to build a
quantum repeater to transmit quantum information across long
distances~\cite{briegel:1998,duan:2001}, as well as to create a
deterministic controlled-\textsc{not} gate between two
non-interacting ions~\cite{duan:2004}. An array of ion traps each
containing a logic ion and an ancilla ion, as illustrated in
Fig.~\ref{fig:photoninter}(b), can be used to perform a remote
deterministic gate.  Here, the logic ions contain quantum
information, while the ancilla ions are used to create a
photon-mediated channel whereby a gate can be performed. These
remote deterministic gates are created by four consecutive steps:
(1) a probabilistic scheme is applied to two ancilla ions until
their entanglement is heralded; (2) local deterministic
controlled-\textsc{not} gates are performed on each logic and
ancilla qubit pairs; (3) the ancilla qubits are measured in an
appropriate basis; (4) single qubit rotations on the logic qubits
are applied based on the results from the measurements of ancilla
ions. The expected number of attempts for a successful entanglement
of the ancilla ions is $\sim~1/P$, depending on the success
probability $P$ of the two-qubit entanglement scheme. Hence, the average
time required to perform a deterministic remote gate is
$\sim~\tau_{rep}/P$.

This scheme can be extended to create a quantum repeater.  Again, we
consider a chain of ion traps spanning the distance, $D$, over which
the information needs to be sent.  In each trap there are two ions,
one to generate a link between each of the adjacent nodes.  For
example, in the $n^\mathrm{th}$ node, photon-mediated entanglement
links one ion to the $(n-1)^\mathrm{th}$ node, and the other ion
links the $(n+1)^\mathrm{th}$ node. A joint Bell state measurement
on the $n^\mathrm{th}$ node leaves the $(n-1)^\mathrm{th}$ node
entangled with the $(n+1)^\mathrm{th}$ node, thereby increasing the
distance between entangled nodes. The time required to generate
entanglement over a distance $D=NL$ with $N$ nodes is thus
$t_{N}=(\tau_{rep}/P)\mathrm{log}_2N$.

\textbf{Generation of Cluster States.} These photon-meditated
quantum networks provide a compelling possibility for operating a
large--scale quantum computer. One universal quantum computation
model is the measurement based cluster state model, which uses a
highly entangled state as an input resource
~\cite{raussendorf:2001,raussendorf:2001a}. This model requires an
initial generation of a large--scale two-dimensional cluster state
and individual single qubit rotations and measurements.

The heralded gate operation discussed earlier (Eq.~\ref{eq:heraldedgate}) is
not unitary gate, which prevents its application to the quantum circuit model.
However, this gate can be used to
generate cluster states because the input qubits for constructing
the cluster state are required to be a superposition state, and the
gate will never give a null result in this case~\cite{duan:2006}.

A 2D square lattice cluster state can be prepared with probabilistic
gates that succeed with probability $P$~\cite{duan:2005}.
Considering a (small) overall failure probability $\epsilon$ for an
$n$-qubit square lattice cluster state, the required time
$T_{cluster}(n,P)$ to generate the state is
\begin{equation}
T_{cluster}(n,P)\simeq
\tau_{rep}\left[\left(\frac{1}{P}\right)^{\mathrm{log}_2(4/P-3)}\,
+\frac{1}{P}\mathrm{log}_2\left(\frac{4[\mathrm{ln}\,(2n/\epsilon)-1]}{P}\right)
+\frac{1}{P}\mathrm{ln}\,(2n/\epsilon)\right],
\label{eq:numberoperation}
\end{equation}
where $\tau_{rep}$ is the operation time for a single attempted gate
operation. From the above equation, the time required to
generate a cluster state is almost independent on the number of
cluster nodes $n$. For example, with $\epsilon = 0.1$ and a two-qubit
gate success probability $P=0.1$
and $\tau_{rep}=1~\hbox{\textmu}$s for the heralded gate discussed
above, we find that the time required for generating the cluster
stare with $n=10^3$ and $n=10^6$ only differs by $0.05\%$. However,
the temporal resource strongly depends on the success probability
$P$. With $P=0.01$, this protocols needs about 5900~years to
generate a 2D square lattice cluster state with $n=10^3$ nodes,
compared with a time of 0.16~second with $P=0.1$. This shows how
critical the improvement of the heralded gate success probability is
for constructing a large scale quantum network.

In this paper, we have described various protocols that rely on
spontaneous emission processes and single photon interference
effects to generate atom--photon and atom--atom entanglement.
Variant schemes, such as generating infrared photonic qubits and
collecting multi-ion emission, have also been studied. We emphasize
that currently the primary issue for realizing a scalable
atom--photon quantum network is the improvement of the probability
of collecting spontaneous emitted photons from trapped ions. To
realize this goal, reflective optics and optical cavities are
suggested for integration into trapped ion systems to enhance the
light collection efficiency. We expect that even a modest
improvement to this efficiency can lead to a large improvement in
the success probability of entangling two distant atomic qubits.
Such improvements also increase the feasibility of deterministic
quantum gate operations between remote atomic qubits, quantum
repeater networks, and cluster state generation.


\begin{acknowledgement}
This work is supported by IARPA under ARO contract, the NSF Physics
at the Information Frontier Program, and the NSF Physics Frontier
Center at JQI. L.L. is supported by JQI postdoctoral fellowship.
\end{acknowledgement}


\providecommand{\WileyBibTextsc}{}
\let\textsc\WileyBibTextsc
\providecommand{\othercit}{} \providecommand{\jr}[1]{#1}
\providecommand{\etal}{~et~al.}

%
%
\end{document}